\documentclass[preprint,showpacs,preprintnumbers,amsmath,amssymb]{revtex4}

\usepackage{graphicx}
\usepackage{dcolumn}
\usepackage{bm}
\usepackage{colordvi}

\begin{document}

\title{Geometry of inertial manifolds probed via a Lyapunov projection method}

\author{Hong-liu Yang$^1$ and G\"unter Radons$^2$}
\affiliation{%
$^1$Institute of Mechatronics,
Reichenhainer Strasse 88, D-09126 Chemnitz, Germany \\
$^2$Institute of Physics,
Chemnitz University of Technology, D-09107 Chemnitz, Germany 
}

\date{\today}
\begin{abstract}
A method for determining the dimension and state space geometry of inertial manifolds of
 dissipative extended dynamical systems is presented. 
 It works by projecting vector differences between reference states 
 and recurrent states onto local linear subspaces spanned by the Lyapunov vectors.
 A sharp characteristic transition of the projection error occurs as soon as 
  the number of basis vectors is increased beyond the inertial manifold dimension. 
  Since the method can be applied using standard orthogonal Lyapunov vectors, 
  it provides a simple way to determine also experimentally inertial manifolds 
and their geometric characteristics.  
\end{abstract}



\maketitle

In ubiquitous natural and laboratory situations dissipative nonlinear partial differential equations (PDEs) are used to model pattern formation, spacetime chaos, turbulence, etc. \cite{stc}. Despite their infinite dimensional nature most dissipative nonlinear PDEs are known to have a finite dimensional attractor owing to the strong dissipation \cite{stc,infinite}. It was further conjectured that apart from some trivial transient decaying process the relevant dynamics of these PDEs takes place on a finite dimensional manifold, named the {\em inertial manifold} (IM)\cite{infinite,inertial}. This concept thus opens up the possibility to model an infinite dimensional PDE system by a finite dimensional dynamical system. The existence of IMs has been proved for a growing list of systems including the Kuramoto-Sivashinsky equation, the complex Ginzburg-Landau equation and some reaction-diffusion equations \cite{inertial-ks,inertial-rd}. Although conceptually important the merit of IMs is largely unexplored due to the complexity of the necessary studying tools \cite{inertial}. 

Motivated by the expectation that infinitesimal perturbations to an IM should somehow capture aspects of finite-size perturbations, a Lyapunov analysis method was applied recently to the study of IMs \cite{ks}: By using covariant Lyapunov vectors \cite{clv} a hyperbolic separation between two sets of Lyapunov vectors was found in dissipative nonlinear PDEs. This splitting became apparant by showing that the angle between the two subspaces spanned respectively by the two sets of vectors is bounded away from zero \cite{ks}. The set of a finite number of mutually entangled Lyapunov vectors associated with positive and weakly negative Lyapunov exponents, named "physical modes", was conjectured to represent the physically relevant dynamics. The linear space spanned by physical modes was thought to serve as the local linear approximation of an IM. The remaining set of vectors corresponding to strongly negative Lyapunov exponents is believed to represent the trivial decaying process to the IM \cite{ks}. In this paper direct support to this conjecture is obtained from the fully nonlinear dynamics via the analysis of projections of differences between recurrent states and the reference state to the subspaces spanned by the Lyapunov vectors. We find below that two competing factors can contribute to the projection error depending on the completeness of the spanned subspace compared to the tangent space of the IM. A sharp transition is observed as the number of spanning Lyapunov modes is increased beyond the IM dimension estimated via Lyapunov analysis \cite{ks}. This finding provides the first direct evidence for the relation between physical modes and the state space geometry of the IM. It confirms the interpretation that the set of physical modes span a linear space approximating the IM locally. Furthermore, the specific variation behavior of the projection error reflects the local quadratic curvature of the IM in general nonlinear systems. It should be emphasized that differences between recurrent states and the reference state used here, are of finite size in contrast to the infinitesimal perturbations considered in a Lyapunov analysis \cite{ks}. Since our projection method can be accomplished also with the standard, orthogonal Lyapunov vectors, one can hope that the existence of IMs, their dimensions, and their curvature properties can be identified easily also from experimental data. 

To demonstrate our finding we use as an example the one-dimensional Kuramoto-Sivashinsky equation \cite{stc,cv}
\begin{equation}
\label{ks}
\partial_t u(x,t) =-\partial^2_xu -\partial^4_xu -u\partial_x u, \qquad x\in[0,L].
\end{equation}
The system is numerically integrated by using the pseudo-spectral scheme and the exponential time differencing method \cite{time-difference} with periodic boundary conditions and a Fourier basis of size $128$. 
The size $L$ is the only control parameter of the system and we use the value $L= 22$ as in \cite{cv} which is sufficiently large to have a structurally stable chaotic attractor and allows also a detailed exploration of the state space geometry of the IM. The standard method of Benettin et al. \cite{benettin} is used to calculate orthogonal Lyapunov vectors (OLVs) $e_n$ and the algorithm of Ginelli et al. \cite{clv} is adopted to get supplementary information from covariant Lyapunov vectors (CLVs). 
 
\begin{figure}
\includegraphics*[scale=0.3]{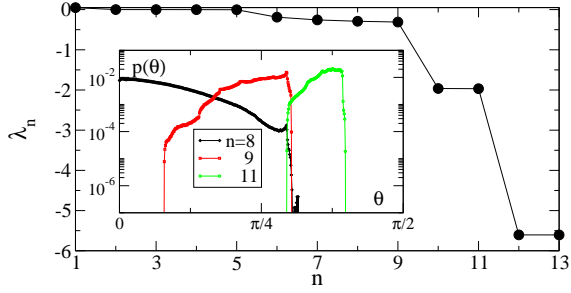}
\caption{\label{fig:f1}
The largest Lyapunov exponents (of $128$) arranged in descending order for the Kuramoto-Sivashinsky equation with $L=22$. The largest exponent has the value $\lambda_1=0.048$ \cite{cv2}. The inset shows the distribution $p(\theta)$ of the manifold angle for several cutting dimensions $n$.
}
\end{figure}

Throughout this paper we assume that the spectrum of Lyapunov exponents is always arranged in descending order and the associated Lyapunov vectors are ordered correspondingly. The values of the Lyapunov exponents shown in Fig. \ref{fig:f1} are in good agreement with Ref.\cite{cv,cv2}.
The angle between two subspaces spanned by the first $n$ CLVs and the remaining ones, respectively, is a fluctuating quantity and is denoted as the manifold angle $\theta$ in the following. It is calculated as in \cite{sub-ang,ks}. Distributions $p(\theta)$ of the manifold angle for several values of $n$ are presented in the inset of Fig. \ref{fig:f1}. It can be seen that the distribution is bounded from zero for $n=9$ and $11$, in contrast to $n=8$ (and lower values), which indicates the (lowest) IM dimension to be $M=9$ for $L=22$ as \cite{cv,cv2}.    

By definition the complete set of Lyapunov vectors span the tangent space of the dynamical system at a given reference state. Analogously, assuming the existence of an IM of dimension $M$, the first $n=M$ Lyapunov vectors span the tangent space of the IM at the same reference state, a local linear approximation of the IM. To quantify the accuracy of the approximation one can define the deviation of the difference vector of reference and recurrent state on the IM from its projection on the spanned linear space as the projection error. In general the IM is a nonlinear manifold and in the neighbourhood of a reference state the quadratic nonlinearity dominates. The projection error for the case with $n=M$ is expected solely due to the nonlinearity/curvature of the IM and would thus decrease quadratically as the reference state is approached. In contrast, a linear space spanned by fewer, say the set of the first $n<M$ Lyapunov vectors can not approximate the IM as well due to its incompleteness compared to the tangent space of the IM. Consequently, the nonzero projections to the missing Lyapunov vectors provide a new source of the projection error and the variation of the projection error with the distance to the reference state will behave differently. Such differences in the projection error by varying the number of spanning Lyapunov vectors can be used to detect the existence of an IM and to estimate its dimension as well. 

To be more specific, to probe the local geometry of the IM at a given reference state $u_0$ we use the collection of recurrent states $u_r$ with the Euclidean distance to the reference state being  $r=\parallel u_r-u_0\parallel$ \cite{distance}. The projection of the state difference vector $\delta u_r=u_r-u_0$ to the linear subspace spanned by the first $n$ Lyapunov vectors can be simply calculated as the sum of projections to each Lyapunov vector individually if OLVs are used \cite{pclv}.
Denoting the length of the projection of the state difference vector $\delta u_r$ on the $m$-th normalized OLV $e_m$ as $p(m,r)=|\delta u_r\cdot e_m|$, the projection error of $\delta u_r$ to the subspace spanned by the first $n$ OLVs can be calculated as $s(n,r)=\sqrt{r^2-\sum^{n}_{m=1} p^2(m,r)}$. Equivalently the projection error can also be obtained as the Euclidean norm of the sum of projections to all OLVs with indices larger than $n$. We study in the following the dependence of the projection error $s(n,r)$ respectively the normalized error $\hat{s}(n,r)=s(n,r)/r$ on the distance $r$ to the reference state and the cutting dimension $n$. $\hat{s}(n,r)$ is simply the projection error obtained from normalized difference vectors $\delta u_r/\parallel \delta u_r \parallel$ and obeys $0 \leq \hat{s}(n,r) \leq 1$.

\begin{figure}
\includegraphics*[scale=0.8]{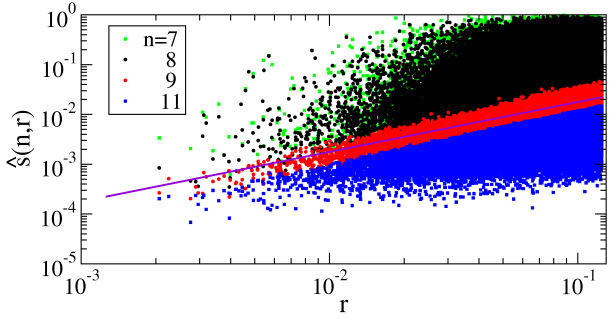}
\caption{\label{fig:f2}
Scatter plots of the normalized projection errors $\hat{s}(n,r)$ for varying distance $r$ between reference and recurrent states and for 
$n=7$, $8$, $9$ and $11$ (superimposed), respectively. A line with slope $1$ is shown to guide the eye.
}
\end{figure}

The variation of the normalized projection error $\hat{s}(n,r)$ with $r$ is shown in Fig. \ref{fig:f2} for several values of $n$. To improve the statistics data from $200$, randomly selected and uncorrelated reference states, each with over $10^4$ recurrent states, are plotted together. A clear difference between the cases with $n < M=9$ and those with $n\geq M$ can be seen in the variation of the upper bound of $\hat{s}(n,r)$ with $r$.
For $n\geq M$ the upper bound of the normalized projection error $\hat{s}$ changes linearly with $r$, which indicates a quadratic dependence of the upper bound of $s(n,r)$ on $r$. This is consistent with the expectation that the projection error for $n\geq M$ is solely due to the nonlinearity of the IM and the quadratic behavior of the upper bound indicates the local quadratic behavior of the IM. For $n <M$ the upper bound of $\hat{s}(n,r)$ is simply the value $1$ independent of the distance $r$ to the reference state, which means that the projection error $s$ can take any value up to size $r$. This is simply due to the nonzero projection of the state difference vector $\delta u_r$ on the subspace spanned by the missing Lyapunov vectors with indices from $n+1$ to $M$. The value $1$ of $\hat{s}(n,r)$ corresponds to the extreme situation that the state difference vector $\delta u_r$ is completely contained in the subspace spanned by the missing Lyapunov vectors. 

\begin{figure}
\includegraphics*[scale=0.3]{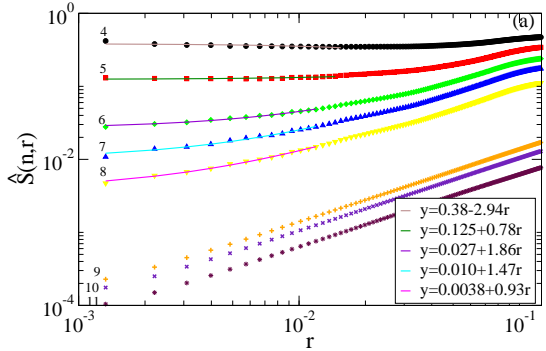}
\caption{\label{fig:f3}
The normalized average projection error $\hat{S}(n,r)$ versus the distance to the reference state $r$ with $n$ from $4$ to $11$ (from top to bottom). Linear fits of the data for $n<M=9$ confirm the saturation of $\hat{S}(n,r)$ to a nonzero value for $r\to 0$, in contrast to the case $n \geq M$. 
}
\end{figure}

We consider now in Fig. \ref{fig:f3} the averaged normalized projection error $\hat{S}(n,r) \equiv \langle \hat{s}(n,r)\rangle$ where $\langle \cdots \rangle$ denotes an average over both the recurrent states and reference states. With decreasing $r$ the mean error $\hat{S}(n,r)$ for $n<M$ shows a clear tendency of saturation to a nonzero value. In contrast, for $n\geq M$ the mean error $\hat{S}(n,r)$ decays continuously in the whole range of $r$ considered. Fitting data to a power law shows that its exponent is almost $1$, which indicates a linear decrease of $\hat{S}(n,r)$ with $r$ for $n\geq M$. For $n<M$ the variation of $\hat{S}(n,r)$ with $r$ in the small-$r$ regime can also be fitted well with a linear function $y=a_0+a_1r$, but now with $a_0 \neq 0$. The nonzero values of $a_0$ confirm the saturation of $\hat{S}(n,r)$ with decreasing $r$. Notice that also here the linear behavior has the same origin as the linear decay of $\hat{S}(n,r)$ for $n \geq M$, namely the local quadratic nonlinearity of the IM. The nonzero saturation value $a_0$ for $n<M$ has a different origin and results from the nonzero projections of state difference vectors to the missing set of OLVs $e_m$ with indices $n< m\leq M$. The value $a_0=\lim_{r\to0} \hat{S}(n,r)$ becoming zero for $n \geq M$ indicates the completeness of the set of the first $n=M$ Lyapunov vectors as a basis of the tangent space of the IM. It can thus be used to detect the existence of an IM and to estimate its dimension $M$.

\begin{figure}
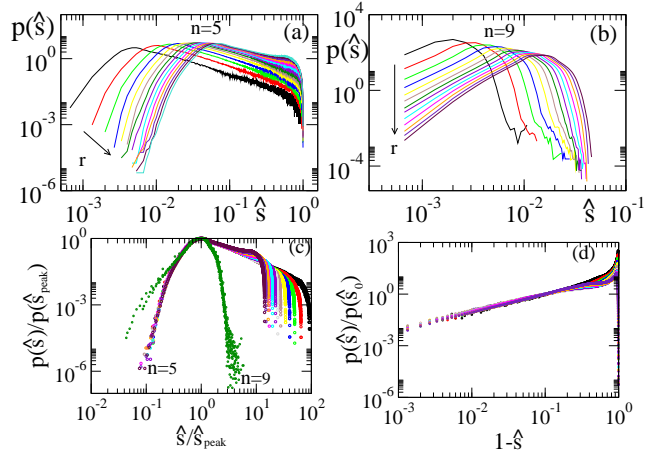

\includegraphics*[scale=0.35]{ks-lv-projection-dist-nc8-9-n2.eps}\\
\includegraphics*[scale=0.3]{ks-lv-projection-dist-normalized-nc-all-near0-08.29-n2.eps}
\includegraphics*[scale=0.3]{ks-lv-projection-dist-normalized-nc-all-near1-08.29-n2.eps}
\caption{\label{fig:f4}
Distributions of the projection error $\hat{s}(n,r)$ for different $r$ (sampled equidistantly in the regime of Fig. \ref{fig:f2}) and two cases (a) $n=5<M$ and (b) $n=9=M$. 
(c) The normalized quantities $p(\hat{s})/p(\hat{s}_{peak})$ versus $\hat{s}(n,r)/\hat{s}_{peak}$ with $n=5$ and $9$. The data collapse for different $r$ in the 
small-$\hat{s}$ regime indicates a common source of the projection errors represented by the small-$\hat{s}$ peak appearing in all cases with different $n$. (d) The normalized quantities $p(\hat{s})/p(\hat{s}_0)$ versus $1-\hat{s}(n,r)$ for $n=5$. The value $s_0$ is arbitarily chosen in the regime close to $1$.
Data collapse to a power law in the regime $\hat{s}(n,r) \simeq 1$ indicates another common source of the projection error for the cases with $n<M$.  
}
\end{figure}

In the following we investigate in more detail the distributions of the normalized projection error $\hat{s}(n,r)$, for which the scatter plot of Fig. \ref{fig:f2} gives already a rough impression. We select two $n$-values, one smaller ($n=5$) than the IM dimension $M=9$ and one equal to $M$. By using these two cases as an example we demonstrate the qualitatively different behavior of the distributions found in general for any $n<M$ and $n \geq M$ by investigating also its $r$-dependence.
As shown in Fig. \ref{fig:f4}, in contrast to the single peak distributions for $n=9$ (panel (b)), the distributions for $n=5<M$ possess on the large-$\hat{s}$ side either a second peak or a shoulder structure (panel (a)). Furthermore, irrespective of the value of $n$ the position $\hat{s}_{peak}$ of the peak on the small-$\hat{s}$ side exhibits a similar linear dependence on $r$, i.e. $\hat{s}_{peak} \sim r$ \cite{yang}. The height of the peak $p(\hat{s}_{peak})$ decreases with $r$ and with increasing $n$ it shows the tendency to become inversely proportional to $r$ \cite{yang}. By using the re-scaled quantities $p(\hat{s})/p(\hat{s}_{peak})$ and $\hat{s}/\hat{s}_{peak}$ distributions for different $r$ values collapse around the small-$s$ peak to a master curve for each $n$ value, which indicates a common source of the projection errors represented by the peak in all cases. For the cases with $n < M$ the rescaled data around the large-$\hat{s}$ peak/shoulder scatter strongly, which implies that the represented projection errors have different origins (see Fig. \ref{fig:f4}(c)). In view of the upper bound of $\hat{s}(n,r)$ for $n<M$ being just $1$, as seen in Fig. \ref{fig:f2}, the rescaled quantities $1-\hat{s}(n,r)$ and $p(\hat{s})/p(\hat{s}_0)$ are used to study the behavior of the distribution near the upper bound (Fig. \ref{fig:f4}(d)). A similar data collapse can indeed be seen for all cases with $n<M$ and the master curves are close to a power law with a $n$-dependent exponent \cite{yang}. 
 Moreover, one can infer from the data collapse with the scaling $1-\hat{s}(n,r)$ that the position of the large-$\hat{s}$ peak/shoulder is independent of $r$. The respective collapse of the projection error distribution data around the two peaks with two different scaling behaviors indicates different sources of the peaks. The small-$\hat{s}$ peak appearing in all distributions represents the projection errors caused by the nonlinearity of the IM. The linear dependence of its peak position $\hat{s}_{peak}$ is consistent with the linear decrease of $\hat{S}(n,r)$ shown in Fig. \ref{fig:f3} for $n\geq M$. The additional large-$\hat{s}$ peak in distributions with $n < M$ represents projection errors due to the inability of the first $n$ Lyapunov vectors to span the tangent space of IM completely. 
The competition of the two sources of errors and the gradual winning of the large-$\hat{s}$ peak explains the linear approach of $\hat{S}(n,r)$ to the saturation value as $r$ goes to zero for $n<M$. 

In summary, we probe the local geometry of IMs by studying the projection error of the difference vectors between recurrent states and the corresponding reference state to the linear subspaces spanned by the first $n$ Lyapunov vectors. The dependence of the projection error on the distance $r$ between the recurrent state and the reference state shows a sharp transition as the number of spanning vector $n$ is increased beyond the IM dimension $M$. The normalized average projection error $\hat{S}(n,r)$ saturates to a nonzero constant for $n<M$ while for $n \geq M$ it decreases linearly to zero as $r$ goes to zero. In contrast to the single peak structure of the probability distribution of the projection error for $n\geq M$ the distribution for $n<M$ has an additional peak at the large-projection-error side. Two sources of the projection errors are identified, the nonlinearity of the IM and the inability of the first $n$ Lyapunov vectors to span the tangent space of the IM completely. And the former becomes the only source of error for $n\geq M$. These changes in projection errors at $n=M$ indicates unambiguously the completeness or incompleteness of the first $n$ Lyapunov vectors as the spanning base vectors of the tangent space of the IM. It supports on one hand the geometric interpretation that the set of physical modes span the tangent space of the IM at the given reference state \cite{ks}. The finite number of physical modes indicates on the other hand the finite dimension of the IM. Some comments are in order now. 1) The Lyapunov projection method introduced here, considers the finite amplitude difference vectors between the recurrent states and reference states, which goes beyond the local stability calculation of the Lyapunov analysis and allows to study the geometric structure of IMs. Our study shows the first direct relation between the state space geometry of the IM and the physical modes identified via Lyapunov analysis. 2) Notice that independent of the manifold angle calculation the IM dimension can be inferred from the transition in the behavior of the projection error. By construction the first $n$ CLVs and OLVs span the same $n$-dimensional linear subspace while the computation of OLVs is much easier\cite{clv,benettin}. The use of OLVs is also preferred due to the ease in the calculation of the projection error. The projection method can thus serve as an {\it independent} fast method to detect IMs and to estimate their dimensions, even from time series of experimental observations \cite{time}. 
3) Compared to the Lyapunov analysis method the projection method probes also nonlocal information of the IM owing to the finite amplitude difference vectors used. It can thus be used for situations where the Lyapunov analysis method may fail to give correct results, for instance due to the fractal folding of the manifold corresponding to physical modes.
4) Similar results were obtained for the complex Ginzburg-Landau equation and we expect the general validity of the reported results for a large class of nonlinear dissipative dynamical systems where the existence of an IM is awaited.

We acknowledge discussions with Hugues Chat\'e, Antonio Politi, Arkady Pikovsky, Kazumasa Takeuchi and financial support from the Deutsche Forschungsgemeinschaft (DFG Grant No. Ra416/6-1).


\begin{thebibliography}{50}
\bibitem{distance}The value $r$ can be normalized by $\sqrt{E} =\lim_{T \to +\infty} (\frac{1}{T L}\int_0^T\mathrm{d} t \int_0^L \mathrm{d} x\frac{u^2(x,t)}{2})^{\frac{1}{2}} $ \cite{cv}. For $L=22$ one has $\sqrt{E} \approx 0.7$.
\bibitem{pclv}With CLVs the projection error can be calculated via a method similar to that used in \cite{sub-ang}.
\end{thebibliography}
\end{document}